\documentclass[twocolumn, journal]{IEEEtran}
\usepackage{graphicx}
\usepackage{amsmath}
\usepackage{amssymb}
\usepackage{color}

\newcommand{\sss}{\scriptscriptstyle}
\newcommand{\non}{\nonumber}
\newcommand{\nid}{\noindent}

\title{Fast Detection of Orthogonal Space-Time Block Codes with Unknown Channel}

\author{Xiaowen Tian, Ming Li,~\IEEEmembership{Member,~IEEE}, Guangyu Ti, and Wenfei Liu
\thanks{Xiaowen Tian, Ming Li, Guangyu Ti, and Wenfei Liu are with the School of Information and Communication Engineering, Dalian University of Technology, Dalian, Liaoning, China, 116024 (e-mail: tianxw@mail.dlut.edu.cn, mli@dlut.edu.cn, tigy@dlut.edu.cn, liuwenfei@dlut.edu.cn).}
\vspace{-0.0 cm}}

\begin{document}

\maketitle

\begin{abstract}
This letter investigates the problem of blind detection of orthogonal space-time block codes (OSTBC) over a quasi-static flat multiple-input multiple-output (MIMO) Rayleigh fading channel. We first introduce a core iterative least-squares (ILS) algorithm to blindly detect OSTBC signals without the knowledge of channel state information (SCI) at the receiver. This ILS algorithm has low computational complexity but may converge to local optimum which offers unreliable detection result. Then, in order to improve the detection performance, we propose an enhanced ILS (E-ILS) approach which is based on statistical analysis of repeated independent ILS procedures on received data. Extensive simulation studies prove the efficiency of the proposed E-ILS algorithm with blind detection performance approaching the optimal maximum-likelihood detector with known CSI.
\end{abstract}

\begin{keywords}
Blind detection, decoding, least-squares, multi-input multi-output (MIMO) communications, orthogonal space-time block code
(OSTBC).
\end{keywords}

\section{Introduction}

In multiple-input-multiple-output (MIMO) communications, space-time coding techniques have emerged as a promising method for effectively utilizing the advantages of multi-antenna diversity. In particular,
orthogonal space-time block codes (OSTBC) have attracted considerable attention in recent literature (for example \cite{Alamouti 98}-\cite{Liang 03} and references therein) because of their maximal diversity gain, simple code construction, and low maximum-likelihood (ML) detection complexity when channel state information (CSI) is available
at the receiver. However, in some scenarios, perfect knowledge of  CSI may be not available at the receiver. For example, CSI might be out-dated due to fast channel fading, or inaccurate if the pilot symbols are under attack. Therefore, the request for blind OSTBC detection with unknown CSI arises.

The blind detection algorithms have been investigated in the past few years.
In \cite{Stoica 03}, \cite{Larsson 03}, a suboptimal algorithm called the cyclic ML was proposed. The idea behind is to decompose the difficult blind detection problem into several simpler subproblems, one of which is coherent ML detection problem. This cyclic ML method can solve the blind detection problem but the accuracy is not satisfactory in some scenarios.
In \cite{Ma 03}-\cite{cui 07}, semidefinite relaxation (SDR) and sphere based blind OSTBC decoding algorithms are proposed. The motivation of this type of approaches is to simplify the blind detection problem to a Boolean quadratic program (BQP) and seek a suboptimal solution to BQP that guarantees polynomial-time worst-case complexity with
respect to the data length. Later on, a novel non-coherent ML OSTBC blind detection algorithm is presented in \cite{dsp 09} which can be performed in polynomial time. While all these algorithms have improved performance of the blind detection with reasonable computational efficiency, there still exists a performance gap between them and the optimal ML decoder with CSI. Therefore, the development of blind detection of OSTBC with both higher performance and lower complexity is needed.

In this paper, we consider the problem of blind OSTBC detection without any knowledge of CSI at the receiver.  Particularly, to simplify the development of the algorithm, we focus our attention on blindly decoding Alamouti code, which is the foundation of OSTBCs. After introducing  signal model in Section \ref{sc:system model}, in Section \ref{sc:Blind Detection Problem} we first develop a core iterative least-squares (ILS) algorithm to blindly detect symbols in Alamouti codes. This ILS algorithm has low computational complexity but may converge to unreliable solution. In an effort to improve detection performance, particularly for small sample size that pose the greatest challenge, in Section \ref{sc:Enhanced algorithm} we propose an algorithmic upgrade referred to as enhanced ILS (E-ILS). This proposed E-ILS algorithm relies on statistical analysis of independent ILS executions on the received data samples. The extensive simulation studies in Section \ref{sc:Simulation}  prove the efficiency of the proposed E-ILS algorithm with blind detection performance approaching the optimal maximum-likelihood detector with known CSI. Finally, a few concluding remarks are drawn in Section \ref{sc:Conclusions}.

\section{System Model}
\label{sc:system model}

To motivate the development of algorithms in this paper, we consider
as an example a MIMO wireless communication system with two transmit antennas and two receive antennas that utilizes the Alamouti OSTBC \cite{Alamouti 98}:
\begin{equation}
\mathbf{C}(n)  =
 \left[ \begin{array}{c c }
  s_{2n-1} & s_{2n} \\
 -s_{2n}^* & s_{2n-1}^* \\
 \end{array}
\right]   \label{eq:encoder matrix}
\end{equation}
where $s_{2n-1}$, $s_{2n} \in \mathcal{A}$  are two symbols in the $n$th Alamouti code block, $\mathcal{A}$ is the normalized constellation (for example, quadrature amplitude modulation (QAM)). If $y_{ij}(n)$ denotes the received signal by the $i$th receive antenna, $i \in \{1,2 \}$, at the $j$th time slot, $j \in \{1,2 \}$, then the received signal of the $n$th  block can be expressed as follows \vspace*{-0.1 cm}
\begin{small}
\begin{equation}
\hspace{-0.1 cm} \left[ \hspace{-0.1 cm} \begin{array}{c c}
  y_{11}(n) & \hspace{-0.2 cm} y_{21}(n) \\
  y_{12}(n) & \hspace{-0.2 cm} y_{22}(n) \\
 \end{array} \hspace{-0.1 cm} \right] \hspace{-0.1 cm} = \hspace{-0.1 cm}
\sqrt{\frac{P}{2}} \mathbf{C}(n) \hspace{-0.1 cm} \left[\hspace{-0.1 cm} \begin{array}{c c}
 h_{11} & \hspace{-0.2 cm} h_{12} \\
 h_{21}  & \hspace{-0.2 cm} h_{22} \\
\end{array} \hspace{-0.1 cm} \right]  \hspace{-0.05 cm} +  \hspace{-0.05 cm} \left[  \hspace{-0.1 cm} \begin{array}{c c}
  n_{11}(n) \hspace{-0.2 cm}& n_{21}(n) \\
  n_{12}(n) \hspace{-0.2 cm} & n_{22}(n) \\
 \end{array} \hspace{-0.1 cm} \right]   \label{eq:received signal form 1} \vspace*{-0.1 cm}
\end{equation}
\end{small}

\noindent where $h_{ij}$ denotes the channel between the $i$th transmit antenna and the $j$th receive antenna, which is assumed to be flat Rayleigh fading; $n_{ij}(n)$ represents additive complex white Gaussian noise  pertinent to the $i$th receive antenna at the $j$th time slot with power $\sigma^2_n$. Due to the special structure of $\mathbf{C}(n)$, the
received signal in (\ref{eq:received signal form 1}) can be rewritten as \vspace*{-0.1 cm}
\begin{small}
\begin{equation}
 \left[ \hspace{-0.1 cm} \begin{array}{c}
  y_{11}(n) \\
  y_{12}^*(n)\\
  y_{21}(n)\\
  y_{22}^*(n)\\
 \end{array} \hspace{-0.1 cm} \right] \hspace{-0.1 cm} = \hspace{-0.1 cm} \sqrt{\frac{P}{2}} \hspace{-0.1 cm} \left[ \begin{array}{c c}
  h_{11} \hspace{-0.2 cm} & h_{21} \\
  h_{21}^* \hspace{-0.2 cm} & -h_{11}^* \\
  h_{12} \hspace{-0.2 cm} & h_{22} \\
  h_{22}^* \hspace{-0.2 cm} & -h_{12}^* \\
 \end{array}\right]       \left[ \begin{array}{c}
  s_{2n-1}   \\
  s_{2n} \\
 \end{array}\right]      \hspace{-0.1 cm}  +  \hspace{-0.1 cm} \left[ \hspace{-0.1 cm} \begin{array}{c}
  n_{11}(n)  \\
  n_{12}^*(n) \\
  n_{21}(n)  \\
  n_{22}^*(n) \\
 \end{array} \hspace{-0.15 cm} \right]
 \label{eq:the n-th snapshot known H} \vspace*{-0.3 cm}
\end{equation}
\end{small}

\vspace*{-0.0 cm}
\nid which can be further expressed in a simpler form as
\begin{equation}
\mathbf{y}(n)=\sqrt{\frac{P}{2}}
\mathbf{H}\mathbf{s}(n)+\mathbf{n}(n).
\label{eq:simply Hs} \vspace*{-0.1 cm}
\end{equation}

\nid It is interesting to note that in (\ref{eq:simply Hs}) the Alamouti structure is embedded in the equivalent channel matrix $\mathbf{H}  $ while the two symbols appear as the elements of a $2 \times 1$ input vector $\mathbf{s}(n) \triangleq [s_{2n-1}, s_{2n}]^T$ and the received signals at time slot 2, $y_{12}(n)$ and $y_{22}(n)$, appear conjugated.

%
%

If the receiver has knowledge of $\mathbf{H}$, optimal ML detection can be adopted to decode the transmitted symbols. However, in some scenarios, the channel $\mathbf{H}$ may be unknown, inaccurate, or out-dated due to fast channel fading. In these cases, an efficient blind detection algorithm is needed to extract symbols $\mathbf{s}(n)$ from $\mathbf{y}(n)$ without known $\mathbf{H}$. To achieve this goal, the receiver collects $N$ blocks/samples and the channel $\mathbf{H}$ remains static over these $N$ samples. Then the received signal   can be formulated  in a matrix form:
\begin{equation}
\mathbf{Y} = \sqrt{\frac{P}{2}}\mathbf{HS}+\mathbf{N}
\label{eq:HS N}
\end{equation}
where $\mathbf{Y}\triangleq\left[\mathbf{y}(1),
\mathbf{y}(2), \ldots ,\mathbf{y}(N)\right] \in \mathbb{C}^{4\times N}$ denotes received signal matrix, $\mathbf{S}\triangleq\left[\mathbf{s}(1),\mathbf{s}(2), \ldots ,
\mathbf{s}(N)\right] \in \mathcal{A}^{2\times N} $ denotes symbol matrix. The goal of this paper is to blindly detect symbol matrix $\mathbf{S}$ from the received signal $\mathbf{Y}$ without known channel matrix $\mathbf{H}$.

Our approach starts with formulating the problem as a joint symbol detection and channel estimation problem with the following least-squares (LS) solution
\begin{equation}
\label{eq:least squares blind}
(\mathbf{\widehat{H}},\mathbf{\widehat{S}}) = \textrm{arg}
\underset{\substack{\mathbf{S}\in \mathcal{A}^{2\times N} \\ \mathbf{H}\in \mathbb{C}^{4\times 2}}}{\textrm{min}} \| \mathbf{Y}-\mathbf{H}\mathbf{S} \|_{\sss F}^2.
\vspace*{-0.1 cm}
\end{equation}
\nid The above LS solution is ML optimal as
long as $\mathbf{N}$ is the white Gaussian noise. The global LS-optimal symbol
matrix $\mathbf{S}$ in (\ref{eq:least squares blind}) can be computed independently of $\mathbf{H}$ by
exhaustive searching over all possible choices under the criterion
function $\|\mathbf{YP}_{\perp S}\|_{\sss F}^2$, i.e.
\begin{equation}
\widehat{\mathbf{S}} = \textrm{arg} \underset{\mathbf{S} \in \mathcal{A}^{2\times N}} {\textrm{min}} \|\mathbf{YP}_{\perp S}\|_{\sss F}^2
\label{eq:search all S} \vspace*{-0.1 cm}
\end{equation}
where $\mathbf{P}_{\perp S} \triangleq \mathbf{I}_2 - \mathbf{S}^H(\mathbf{SS}^H)^{-1}\mathbf{S}$. Exhaustive searching has complexity of exponential in $2N$ (total size of symbols). Considering this unacceptable computational cost, we attempt to obtain a quality approximation of the solution of (\ref{eq:least squares blind}) by alternating least-squares estimation of channel $\mathbf{H}$ and detection of symbol $\mathbf{S}$, iteratively, as described in the next section.

%

\section{Iterative least-squares Algorithm}
\label{sc:Blind Detection Problem}

Assuming channel $\mathbf{H}$ is known, then the least-squares detection of symbol matrix $\mathbf{S}$ can be obtained by
\begin{equation}
\widehat{\mathbf{S}} = \textrm{arg}
\underset{ \mathbf{S} \in \mathcal{A}^{2\times N}}
{\textrm{min}}
\| \mathbf{Y}-\mathbf{H}\mathbf{S} \|_{\sss F}^2.
\label{eq:least squares known H}
\end{equation}

\noindent While the least-squares
estimate of $\textbf{S}$ \textit{over complex field} is
\begin{eqnarray}\label{eq:ILS2}
\widehat{\mathbf{S}}^{\mathrm{complex}} &
= &  \textrm{arg}
\underset{ \mathbf{S} \in \mathbb{C}^{2\times N}}
{\textrm{min}}
\| \mathbf{Y}-\mathbf{H}\mathbf{S} \|_{\sss F}^2  \non \\
& = &(\mathbf{H}^H\mathbf{H})^{-1} \mathbf{H}^H \mathbf{Y},
\end{eqnarray}

\noindent we suggest the approximate digital (finite modulation alphabets) solution as
\begin{eqnarray}
\widehat{\mathbf{S}}^{\mathrm{digital}} &
= &  \textrm{arg}
\underset{ \mathbf{S} \in \mathcal{A}^{2\times N}}
{\textrm{min}}
\| \mathbf{Y}-\mathbf{H}\mathbf{S} \|_{\sss F}^2  \non \\
& \approx & \mathcal{U}\{(\mathbf{H}^H\mathbf{H})^{-1} \mathbf{H}^H \mathbf{Y}\} \label{eq:known H detection S}
\end{eqnarray}

\nid where $\mathcal{U}\{\cdot\}$ denotes the projection of the complex value into the closest constellation point.

Assuming, in return, that the symbol matrix $\mathbf{S}$ is known, we attempt to estimate channel $\mathbf{H}$. Since $\mathbf{H}$ has correlated elements (e.g. $h_{1,1}$ and $-h_{1,1}^*$, see (\ref{eq:the n-th snapshot known H})), directly estimating $\mathbf{H}$ from $\mathbf{Y}$ with known $\mathbf{S}$ is not appropriate. To facilitate the algorithm development, we recall the original received signal form (\ref{eq:received signal form 1}) and rewrite it in an matrix representation:
 \begin{equation}
\widetilde{\mathbf{Y}}(n)=
\sqrt{\frac{P}{2}}\mathbf{C}(n)\mathbf{G}+\mathbf{Z}(n)  \label{eq:CG form n}
\end{equation}
\nid where $\widetilde{\mathbf{Y}}(n) \triangleq \left[ \begin{array}{c c}
  y_{11}(n) & y_{21}(n) \\
  y_{12}(n) & y_{22}(n) \\
 \end{array} \right]$,  $\mathbf{G} \triangleq
\left[ \begin{array}{c c}
 h_{11} &h_{12} \\
 h_{21}  &h_{22} \\
\end{array} \right]$, $\mathbf{Z}(n)$ is noise term. If $N$ samples are obtained, we can stack the received signals as
\begin{equation}
 \widetilde{\mathbf{Y}} = \sqrt{\frac{P}{2}}\mathbf{CG}+\mathbf{Z}.
 \label{eq:CG form N samples}
\end{equation}

\noindent where $\widetilde{\mathbf{Y}}\triangleq\left[
\widetilde{\mathbf{Y}}(1)^T,\widetilde{\mathbf{Y}}(2)^T, \ldots,
\widetilde{\mathbf{Y}}(N)^T \right]^T  \in \mathbb{C}^{2N\times2}$ is the stacked signal,
$\mathbf{C}\triangleq\left[\mathbf{C}(1)^T,\mathbf{C}(2)^T,\ldots,\mathbf{C}(N)^T \right]^T \in \mathcal{A}^{2N\times2}$ contains $N$ transmitted Alamouti codes in which the symbols are embedded. If the symbol matrix $\mathbf{S}$ is known, the Alamouti codes $\mathbf{C}$ can be constructed by (\ref{eq:encoder matrix}) and the least-squares solution of channel matrix $\mathbf{G}$ is given by
\begin{eqnarray}
\widehat{\mathbf{G}} & = & \textrm{arg}
\underset{ \mathbf{G} \in \mathbb{C}^{2\times2}}
{\textrm{min}}
\| \widetilde{\mathbf{Y}}-\mathbf{C}\mathbf{G} \|_{\sss F}^2 \non \\  & = &(\mathbf{C}^H\mathbf{C})^{-1}\mathbf{C}^H \widetilde{\mathbf{Y}}.
\label{eq:least squares known C}
\end{eqnarray}

The \textit{iterative least-squares} (ILS) algorithm motivated by (\ref{eq:known H detection S}) and (\ref{eq:least squares known C}) is now straightforward. Arbitrarily initialize symbol matrix $\mathbf{S}$, construct the stacked Alamouti code $\mathbf{C}$ with this $\mathbf{S}$,, and estimate channel matrix $\mathbf{\widehat{G}}$ by (\ref{eq:least squares known C}). With the estimated channel $\mathbf{\widehat{G}}$, construct an equivalent channel matrix $\mathbf{H}$ used in (\ref{eq:HS N}) and detect symbol matrix $\mathbf{\widehat{S}}$ by (\ref{eq:known H detection S}). Alternate the calculation between (\ref{eq:known H detection S}) and (\ref{eq:least squares known C}) iteratively until convergence is found. This ILS algorithm is summarized in Table \ref{tb:ILS} where superscript $d$ denotes iteration index. For $2\times2$ Alamouti code considered in this paper, the computational complexity of each iteration of the ILS algorithm is $\mathcal{O}(N)$ and, experimentally, the number of iterations is between 2 and 5 in general.

\begin{center}
\begin{table}[t]  \vspace{0.0 cm}\caption{
Iterative  least-squares (ILS) Algorithm} \vspace{-0.3cm}
\begin{center}
\begin{tabular}{l}
\hline \hline \vspace{-0.2 cm}\\
 \hspace{-0.2 cm} \textbf{Step 1.} \hspace{0.2 cm} $d=0$; \\
\hspace{1.0 cm} Initialize $\mathbf{S}^{(0)} \in \mathcal{A}^{2 \times N}$ arbitrarily.\\
\hspace{-0.2 cm} \textbf{Step 2.} \hspace{0.2 cm} $d=d+1$;\\
\hspace{1.0 cm}  Construct $\mathbf{C}^{(d)}$ using $\mathbf{{\widehat{S}}}^{(d-1)}$;\\
\hspace{1.0 cm}\begin{footnotesize}$\mathbf{{\widehat{\mathbf{G}}}}^{(d)}
= \left[(\mathbf{C}^{(d)})^H\mathbf{C}^{(d)} \right]^{-1} (\mathbf{C}^{(d)})^H\mathbf{\widetilde{Y}}$;
\end{footnotesize} \\
\hspace{1.0 cm} Construct $\mathbf{H}^{(d)}$ using $\mathbf{{\widehat{G}}}^{(d)}$;\\
\hspace{0.9 cm} \begin{footnotesize}
$\mathbf{{\widehat{S}}}^{(d)}= \left[ (\mathbf{H}^{(d)})^H\mathbf{H}^{(d)} \right]^{-1}(\mathbf{H}^{(d)})^{H}
\mathbf{Y}$.
\end{footnotesize} \\
\hspace{-0.2 cm} \textbf{Step 3.} \hspace{0.2 cm} Repeat Step 2 until
$\mathbf{{\widehat{S}}}^{(d)} =
\mathbf{{\widehat{S}}}^{(d-1)}$.\\
\hline
\end{tabular}\label{tb:ILS}\vspace{-0.2 cm}
\end{center}
\end{table}
\end{center}

\section{Enhanced ILS Algorithm}
\label{sc:Enhanced algorithm}

The reliability of the ILS convergence point depends heavily on the initialization. With arbitrary initialization, convergence of the ILS algorithm described in Table \ref{tb:ILS} to the optimal (least-squares  fit) solution of (\ref{eq:least squares blind}) is not always assured. To that respect, re-execution of the ILS algorithm with distinct re-initialization is a promising solution to avoid those unreliable convergence point.

\begin{center}
\begin{figure}[t]
\begin{center}
\hspace{0.0 cm}
  \includegraphics[width=2.4 in ]{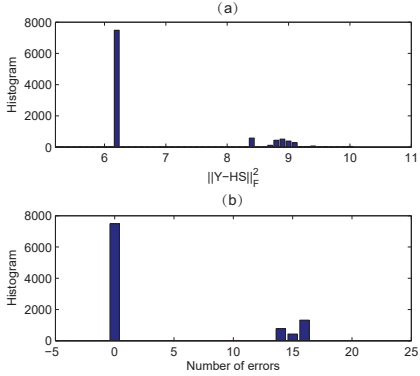}
  \vspace{-0.5 cm}
  \caption{Experiment with ILS of Table \ref{tb:ILS}:
 (a) Histogram of the LS residual, (b) histogram of the number of errors
  ($N=20$, $P=8$dB, $\sigma_n^2 =1 $). }\label{fig:corr}\vspace{-0.4 cm}
\end{center}
\end{figure}
\end{center}

Our proposal is motivated by following experiment. Based on $N=20$ samples with $P = 6$dB transmit power and $\sigma^2_n=1$ noise power, we execute the ILS algorithm $10,000$ times with distinct arbitrary initialization of each. With the detected symbols $\mathbf{\widehat{S}}$ and
estimated channel $\widehat{\mathbf{H}}$ returned by
each ILS algorithm upon convergence, we plot the histograms of LS residual $\| \mathbf{Y} - \widehat{\mathbf{H}}\widehat{\mathbf{S}} \|_{\sss F}^2$ and the
numbers of error symbols in $\mathbf{\widehat{S}}$ in Fig. \ref{fig:corr}(a)
and (b), respectively. The results in Fig. \ref{fig:corr} reveal that most of the ILS convergence points
are reliable (close to minimal LS fit and having few or no error symbols)
and those unreliable ILS convergence points result in much higher
LS residual. \vspace{-0.3 cm}

\vspace{-0.0 cm}

Motivated by this observation, we first propose to re-execute
ILS algorithm $Q$ times with distinct arbitrary initialization and obtain  $Q$ returned ILS solutions, which are denoted as $\mathcal{M}\triangleq$ $\{(\widehat{\mathbf{H}}_1,\widehat{\mathbf{S}}_1), \ldots, (\widehat{\mathbf{H}}_Q,\widehat{\mathbf{S}}_Q)\}$. To assess which of these $Q$ returned ILS solutions in $\mathcal{M}$  has superior least-squares fit, we simply feed $(\widehat{\mathbf{H}}_q,
\widehat{\mathbf{S}}_q)$ to LS residual calculation $R_q \triangleq \| \mathbf{Y} -
\widehat{\mathbf{H}}_q\widehat{\mathbf{S}}_q \|_{\sss F}^2$, $q=1,\ldots, Q$, and choose the best pair $(\widehat{\mathbf{H}}^{*},
\widehat{\mathbf{S}}^{*})$ as \vspace{-0.1 cm}
\begin{equation}
(\widehat{\mathbf{H}}^{*},
\widehat{\mathbf{S}}^{*}) = \textrm{arg}
\underset{ (\widehat{\mathbf{H}},\widehat{\mathbf{S}}) \in \mathcal{M}}
{\textrm{min}}
\| \mathbf{Y}-\widehat{\mathbf{H}}\widehat{\mathbf{S}} \|_{\sss F}^2. \label{eq:LS minimum} \vspace{-0.1 cm}
\end{equation}

This straightforward approach provides high probability of finding an LS optimal solution from redundant ILS solutions. Larger $Q$ can potentially provide better performance but cause higher computational complexity. In an effort to reduce the complexity, we attempt to reduce ILS executions without significant performance loss. The experiment results in Fig. \ref{fig:corr} indicate that reliable ILS convergence points not only have minimum LS residual, but also are majority. Motivated by this finding, in stead of seeking the reliable detection when finishing all $Q$ ILS executions, we
propose to evaluate the residual value $R_q$ after each ILS execution. Specifically, ILS re-execution will be stopped if
1) $R_q$ obtained by current ILS execution is \textit{minimum} (i.e. $R_{q} \leq R_{i}, i\in\{ 1, \ldots, q-1\}$) and also \textit{majority} (i.e. having $T$ same values in previous ones $\{R_1, \ldots, R_{q-1}\}$),
or 2) the ILS is carried out $Q$ times.
For condition 1, once $R_q$ satisfies both minimum and majority criterion,
the current detection is considered reliable and complexity is reduced due to less ILS executions. For condition 2, we turn to find the best detection from all $Q$ results based on only minimum criteria as (\ref{eq:LS minimum}).
The majority threshold $T$ is typically selected as $T \in [2-6]$ with which the ILS executions can be significantly reduced. The details of proposed E-ILS algorithm are described in Table \ref{tb:Trade-off ILS}.


%

\begin{center}
\begin{table}[t]  \vspace{-0.0 cm}
\caption{Enhanced Iterative least-squares (E-ILS) Algorithm}
\vspace{-0.3 cm}
\begin{center}
\begin{tabular}{l}
\hline \hline \vspace{-0.2 cm}\\
\textbf{Input:} \hspace{0.2 cm}  $Y$, $Q$, $T$.\\
\textbf{Initialization:} \hspace{0.2 cm}  $q=0$, $R_{min} = \mathrm{inf}$, \texttt{run\_flag}$ = 1$.\\
\textbf{While } \hspace{0.2 cm} $q \leq Q$ and \texttt{run\_flag}$  = 1$ \\
\hspace{0.3 cm}  $q=q+1$;\\
\hspace{0.3 cm} Execute ILS to obtain $(\widehat{\mathbf{H}}_q, \widehat{\mathbf{S}}_q)$ and calculate LS residual $R_q$; \\
\hspace{0.3 cm}  \textbf{If} $R_q \leq R_{min}$ \\
\hspace{0.7 cm}              $R_{min} = R_q$; \\
\hspace{0.7 cm}              $\widehat{\mathbf{S}}^*  = \widehat{\mathbf{S}}_q$; \\
\hspace{0.9 cm}   \textbf{If} $R_q$ has $T$ same values in $\{R_1, \ldots, R_{q-1}\}$\\
\hspace{1.2 cm} \texttt{run\_flag}$ = 0$;\\
\hspace{0.9 cm}  \textbf{Endif}  \\
\hspace{0.3 cm}  \textbf{Endif}  \\
 \textbf{End while}  \\
\textbf{Output:} \hspace{0.2 cm}  $ \widehat{\mathbf{S}}^*$.\\
\hline\vspace{-0.7 cm}
\end{tabular}\label{tb:Trade-off ILS}\vspace{-0.0 cm}
\end{center}
\end{table}
\end{center}

 \vspace{-0.5 cm}

\section{Simulation Studies}
\label{sc:Simulation}

\begin{center}
\begin{figure}[t]
\begin{center}
\vspace{-0.0 cm}
  \includegraphics[width=2.90 in ]{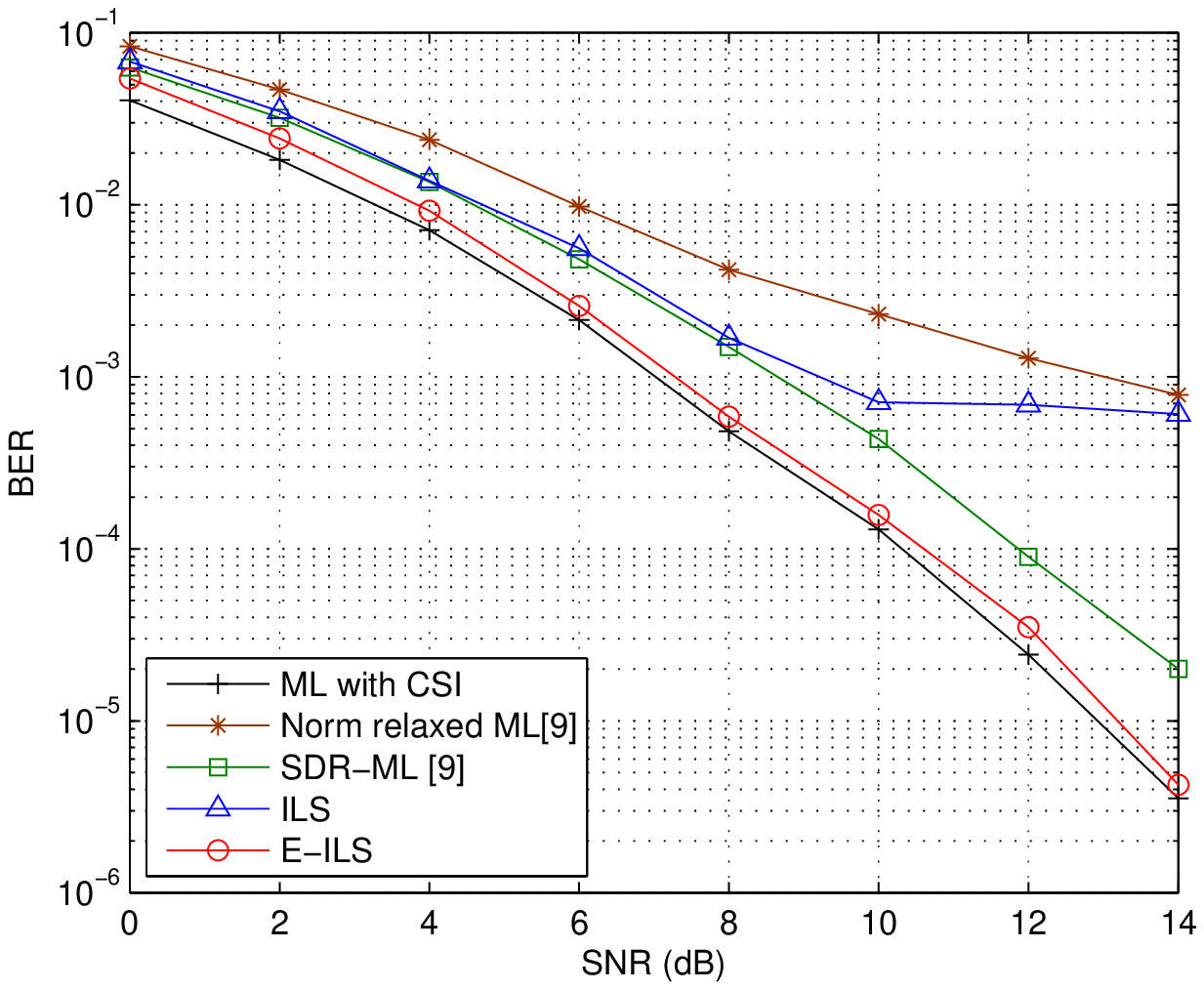}
  \vspace{-0.3 cm}
  \caption{BER versus SNR (BPSK, $N=20$, $D=20$, $T=2$). }\label{fig:BPSK_four_al_Nr2}\vspace{-0.0 cm}
\hspace{0.0 cm}
\includegraphics[width=2.90 in ]{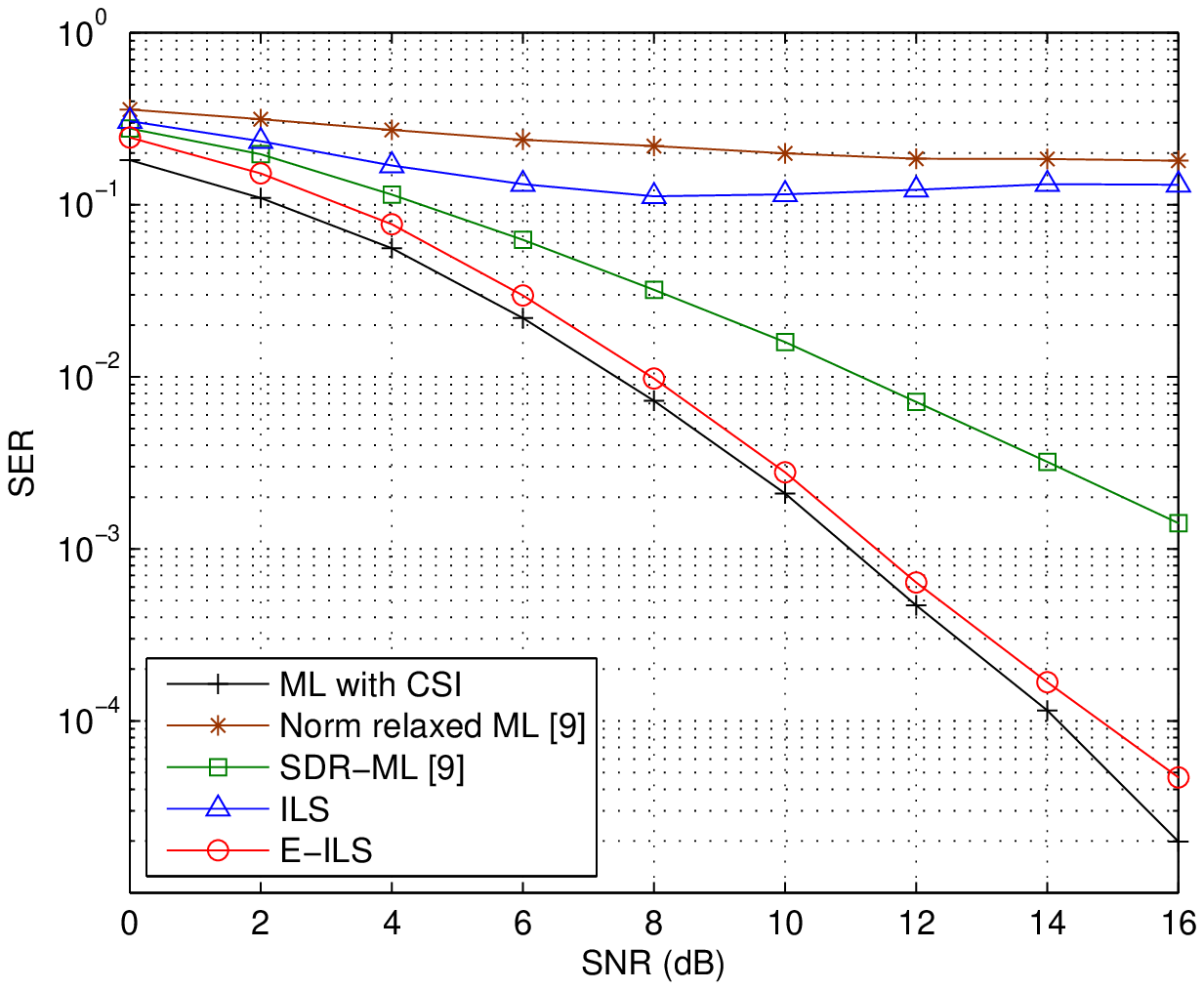}
\vspace{-0.3 cm}
\caption{SER versus SNR (QPSK, $N=20$, $Q=20$, $T=4$). }\label{fig:QPSK_four_al_Nr2}\vspace{-0.2 cm}
\end{center}
\end{figure}
\end{center}

In the following, we present extensive simulation studies to illustrate the performances of the proposed ILS and E-ILS algorithms. As a reference, ideal ML detection with perfect CSI is also included. For comparison purpose, we also evaluate two well-known blind detection algorithms, SDR-ML and norm relaxed ML blind detections, presented in \cite{Ma 06}. We first consider BPSK modulation with the number of samples fixed at $N=20$. For the E-ILS algorithm, we set the maximum number of ILS executions as $Q=20$ and the majority threshold as $T = 2$. The bit-error-rate (BER) of each algorithm is plotted in Fig. \ref{fig:BPSK_four_al_Nr2} as a function of signal-to-noise-ratio (SNR) which is defined as $\mathrm{SNR}=P/\sigma_n^2$. \vspace{-0.3 cm}

It can be observed from Fig. \ref{fig:BPSK_four_al_Nr2} that the proposed E-ILS algorithm outperforms its competitors and has very impressive performance close to the optimal ML detection with perfect CSI. In this experiment, the average number of iterations in each ILS is $2.9345$ and the average number of ILS executions in E-ILS is $3.4741$, which verify that E-ILS has very low complexity. In Fig. \ref{fig:QPSK_four_al_Nr2}, the same simulation is repeated with QPSK modulation and similar conclusions can be drawn.

\vspace{-0.3 cm}


\begin{center}
\begin{figure}[t]
\begin{center}
\hspace{0.0 cm}
  \includegraphics[width=2.90 in ]{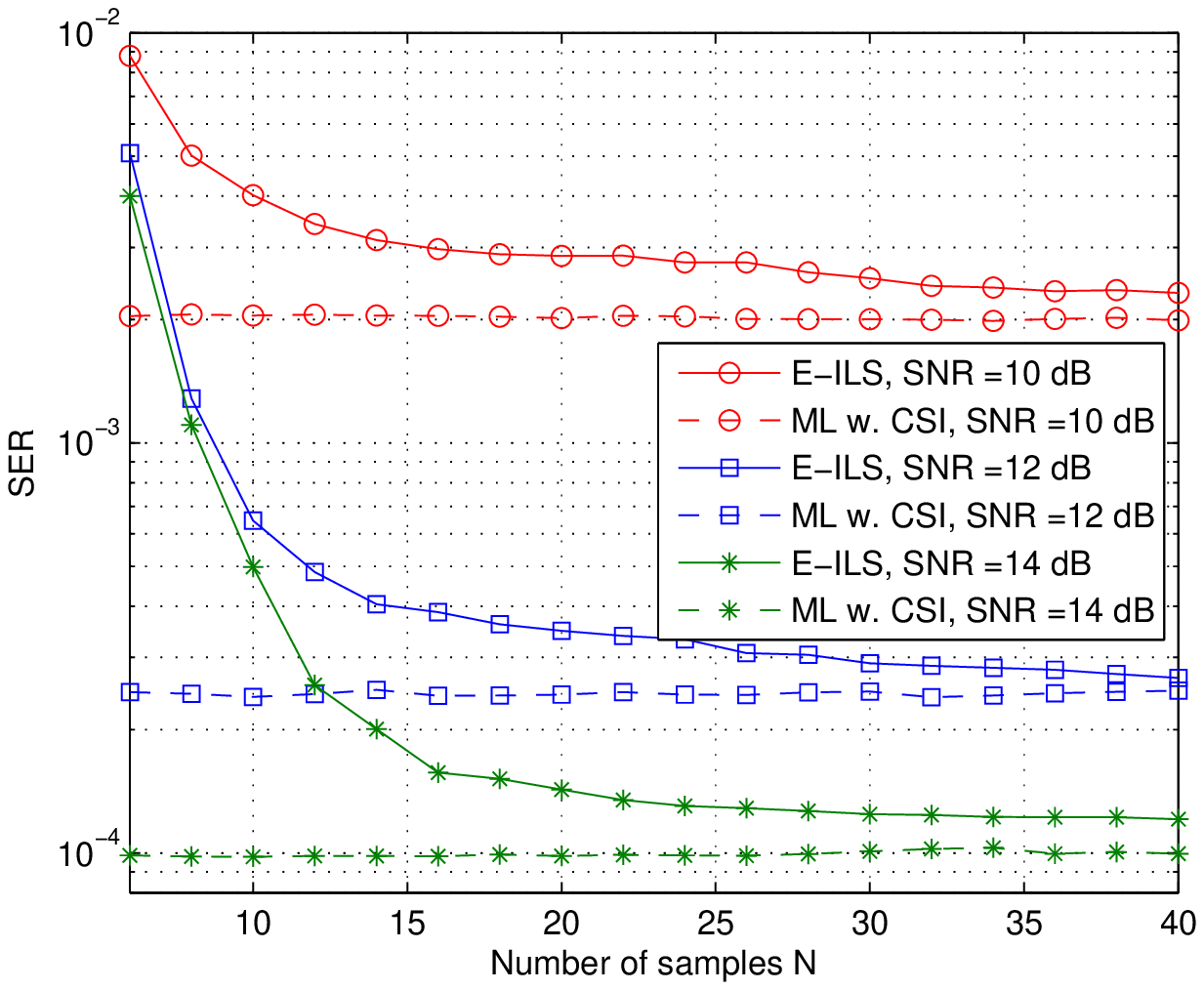}
  \vspace{-0.3 cm}
  \caption{BER versus number of samples $N$ (QPSK, $Q=20$, $T=4$). }\label{fig:QPSK 2_D}\vspace{0.0 cm}
  \includegraphics[width=2.90 in ]{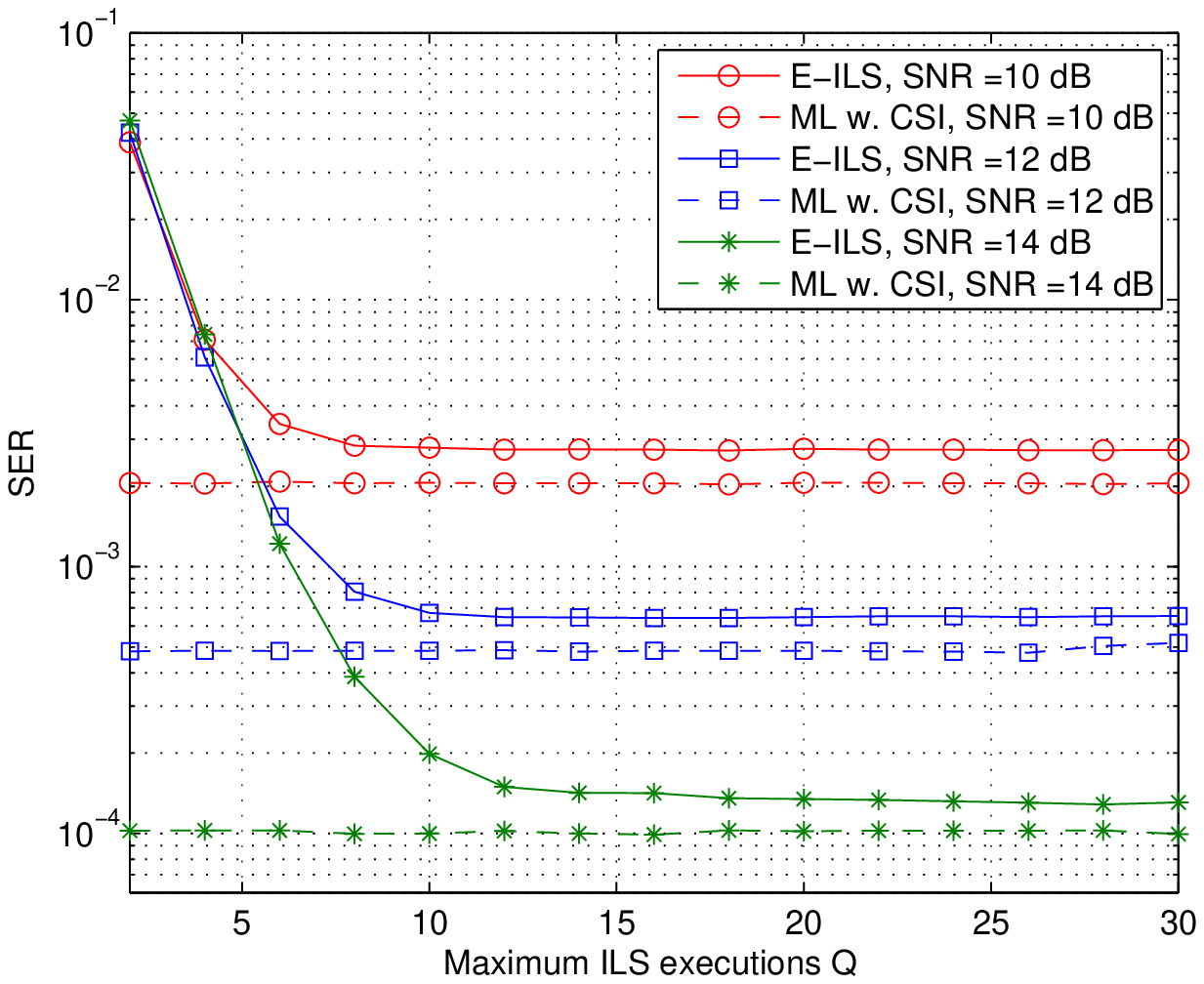}
  \vspace{-0.3 cm}
  \caption{SER versus maximum number of ILS executions $Q$ (QPSK, $N=20$, $T = min\{4,Q-1\}$).}\label{fig:QPSK 2_P}\vspace{-0.2 cm}
\end{center}
\end{figure}
\end{center}

The dependence of error probability on the size of the data samples $N$ is illustrated in Fig. \ref{fig:QPSK 2_D} with SNR fixed at 10dB, 12dB, and 14dB, respectively.  The findings corroborate the conclusions drawn from Fig. \ref{fig:QPSK_four_al_Nr2} and verify that E-ILS has good performance with small sample size. Finally, in Fig. \ref{fig:QPSK 2_P} we illustrate the error probability as a function of the maximum number of ILS executions $Q$. The results indicate that a small value of $Q$ (for example $Q = 20$ in our simulation) can provide excellent trade-off between performance and complexity.


\section{Conclusions}
\label{sc:Conclusions}

In this paper, we considered the problem of blind detection of OSTBC  without the knowledge of CSI at the receiver. A core iterative least-squares algorithm was first presented to blindly detect symbol information. To achieve satisfactory performance, we also proposed an enhanced ILS (E-ILS) algorithm which is based on statistical analysis of repeated independent ILS processing on the received data. Simulation studies demonstrated that the proposed E-ILS algorithm has effective blind detection performance with probability of error close to optimal maximum-likelihood decoder with known CSI.


\begin{thebibliography}{99}

\bibitem{Alamouti 98} S. M. Alamouti, ``A simple transmit diversity technique for wireless communications,'' \textit{IEEE J. Sel. Areas Commun.}, vol. 16, no. 8, pp. 1451-1458, Oct. 1998.

\bibitem{Tarokh 99} V. Tarokh, H. Jafarkhani, and A. R. Calderbank, ``Space-time block codes from orthogonal designs,'' \textit{IEEE Trans. Inf. Theory}, vol. 45, no. 5, pp. 1456-1467, July 1999.

\bibitem{Stoica 01} P. Stoica and G. Ganesan, ``Space-time block codes: A maximum SNR approach,'' \textit{IEEE Trans. Inf. Theory}, vol. 47, no. 4, pp. 1650-1656, May 2001.

\bibitem{Su 03} W. Su and X.-G. Xia, ``On space-time codes from complex orthogonal designs,'' \textit{Wireless Pers. Commun.}, vol. 25, pp. 1-26, April 2003.

\bibitem{Liang 03} X.-B. Liang, ``Orthogonal designs with maximal rates,'' \textit{IEEE Trans. Inf. Theory}, vol. 49, no. 10, pp. 2468-2503, Oct. 2003.


\bibitem{Stoica 03} P. Stoica and G. Ganesan, ``Space-time block codes: Trained, blind, and semi-blind detection,'' \textit{Digital Signal Process.}, vol. 13, no. 1, pp. 93-105, Jan. 2003.

\bibitem{Larsson 03} E. G. Larsson, P. Stoica, and J. Li, ``Orthogonal space-time block codes: Maximum likelihood detection for unknown channels and unstructured interferences,'' \textit{IEEE Trans. Signal Process.}, vol. 51, no. 2, pp. 362-372, Feb. 2003.



\bibitem{Ma 03} W.-K. Ma, P. C. Ching, T. N. Davidson, and X.-G. Xia, ``Blind maximum-likelihood decoding for orthogonal space-time block codes: A semidefinite relaxation approach,'' in \textit{Proc. IEEE Global Commun. Conf. (Globecom)}, San Francisco, CA, Dec. 2003, vol. 4, pp. 2094-2098.

\bibitem{Ma 06} W.-K. Ma, B.-N. Vo, T. N. Davidson, and P.-C. Ching, ``Blind ML detection of orthogonal space-time block codes: Efficient high-performance implementations,'' \textit{IEEE Trans. Signal Process.}, vol. 54, no. 2, pp. 738-751, Feb. 2006.

\bibitem{cui 07} T. Cui, and C. Tellambura, ``Efficient blind receiver design for orthogonal space-time block codes,'' \textit{IEEE Trans. Wireless Commun.}, vol. 6, no. 5, May 2007.

\bibitem{dsp 09} D. S. Papailiopoulos, and G. N. Karystinos, ``Optimal OSTBC sequence detection over unknown correlated fading channels,'' in \textit{Proc. Asilomar Conf.}, Pacific Grove, CA, Nov, 2009, pp. 1441-1445.

%
%
%

\end{thebibliography}
\end{document}